\documentclass[twocolumn,aps,superscriptaddress,showpacs,nofootinbib,floatfix]{revtex4}
\usepackage{epsfig,bm,feynmf}
\usepackage{graphicx}
\usepackage{amsmath}
\usepackage{dcolumn}

\usepackage{hyperref}
\hypersetup{
  colorlinks,
  unicode=True,
  linkcolor=Blue,
  citecolor=Blue,
  urlcolor=Blue,
}

\usepackage{graphicx}
\usepackage[usenames,dvipsnames,svgnames,table]{xcolor}

\begin{document}


\title{Heavy quark mass near the phase transition}

\author{Taesoo Song\footnote{corresponding author}}\email{t.song@gsi.de}
\affiliation{GSI Helmholtzzentrum f\"{u}r Schwerionenforschung GmbH, Planckstrasse 1, 64291 Darmstadt, Germany}

\author{Qi Zhou}\email{qizhou@mails.ccnu.edu.cn}
\affiliation{Key Laboratory of Quark \& Lepton Physics (MOE) and Institute of Particle Physics,
Central China Normal University, Wuhan 430079, China}
\affiliation{Institut f\"ur Theoretische Physik, Johann Wolfgang Goethe-Universit\"at,Max-von-Laue-Str.\ 1, D-60438 Frankfurt am Main, Germany}


\begin{abstract}
Assuming that the number densities of heavy flavor in hadron gas and in QGP are same at $T_c$, we obtain the effective mass of heavy quark at $T_c$ from the comparison with the hadron resonance gas model which well describes particle yield in heavy-ion collisions.
We find that charm quark mass at vanishing baryon chemical potential is around 1.8 GeV which is much heavier than QCD bare mass and close to $D$ meson mass. The mass slightly increases with increasing baryon chemical potential and then decreases.
On the other hand, anticharm quark mass constantly decreases with increasing baryon chemical potential.
Bottom quark mass has a similar pattern.
Extending the hadron resonance gas model to a bit higher temperature beyond $T_c$, the effective masses of charm and bottom quarks decrease with increasing temperature. 
\end{abstract}


\maketitle

\section{Introduction}
Heavy flavor is one of the promising probe particles to search for the properties of a hot and dense matter produced in relativistic heavy-ion collisions~\cite{STAR:2014wif,STAR:2018zdy,CMS:2017qjw,ALICE:2021rxa,Gossiaux:2009mk,He:2011qa,Uphoff:2012gb,Cao:2013ita,Cassing:1997kw,Song:2015sfa,Plumari:2017ntm,Cao:2019iqs,Beraudo:2022dpz,Cao:2018ews,Rapp:2018qla,Xu:2018gux,Zhao:2023nrz}.

The properties of a partonic matter, for example, the equation-of-state (EoS) are provided in lattice QCD, not only at zero baryon chemical potential but also at finite chemical potential to some extent~\cite{Borsanyi:2010cj,Borsanyi:2012cr}.
The macroscopic properties of quark-gluon plasma (QGP) from the lattice calculations can be interpreted as the gathering of quasiparticles in microscopic view~\cite{Gorenstein:1995vm,Levai:1997yx,Peshier:1995ty,Peshier:2005pp,Plumari:2011mk}.
Since QCD matter is strongly interacting especially near the critical temperature for the phase transition, more reasonable description will be the dynamical quasiparticle model (DQPM) which takes into account not only the dressing of parton in QGP but also the spectral width contributed from interactions in matter~\cite{Cassing:2007nb,Moreau:2019vhw}.

Different from light flavors, the effective mass of heavy flavor cannot be extracted from the EoS, because its contribution is too little.
Heavy quark mass in QGP will be different from the bare mass in QCD Lagrangian, because it is dressed in the partonic matter.
Recently thermal production of charm quark pair in relativistic heavy-ion collisions was studied and it was found that the experimental data favors the production of massive charm quarks rather than charm quarks with the bare mass~\cite{Song:2024hvv}.

Heavy quark mass in QGP is important, because it is one of necessary ingredients to study quarkonium in QGP.
For example, the properties of quarkonium in QGP are obtained in many phenomenological models~\cite{Grandchamp:2003uw,Yan:2006ve,Song:2011xi,Song:2011nu,Strickland:2011aa,Brambilla:2016wgg,Andronic:2024oxz} by solving the Shr\"{o}dinger equation with a heavy quark potential which is provided by lattice QCD~\cite{Kaczmarek:2002mc,Satz:2005hx,Lafferty:2019jpr}.
One of the important parameters in the Shr\"{o}dinger equation is heavy quark mass.
It is also needed to define the binding energy of quarkonium which is the difference between twice heavy quark mass and quarknium mass in QGP.

The hadron resonance model (HRG) assumes that the hadron gas phase is composed of all physical states of hadrons including both ground and excited states in grand canonical ensemble~\cite{Cleymans:1999st,Becattini:2000jw,Braun-Munzinger:2001hwo}.
The HRG model well describes the multiplicities not only of light particles but also of heavy particles such as charmonium, taking into account charm fugacity~\cite{Andronic:2019wva}, in relativistic heavy-ion collisions, 

According to lattice calculations~\cite{Borsanyi:2010cj,Borsanyi:2012cr} the phase transition between QGP and hadron gas phase is crossover at low baryon chemical potential.
Therefore all physical quantities on both sides will be smoothly connected around $T_c$.
In fact, thermal quantities of QCD matter such as energy density, pressure and entropy density show smooth transitions from hadron gas to QGP in the lattice calculations~\cite{Karsch:2003vd,Borsanyi:2010cj}.
Since all physical quantities smoothly change during the phase transition, (anti)charm number density 
will also be continuous.
In other words, sudden annihilation or creation of charm quark pairs will not take place.
Even though the dissociation temperature of charmonium is assumed around $T_c$, its dissociation and production will simultaneously happen through the detailed balance and it won't drastically change the number density of (anti)charm quark.

\begin{figure} [h!]
\includegraphics[width=8.5 cm]{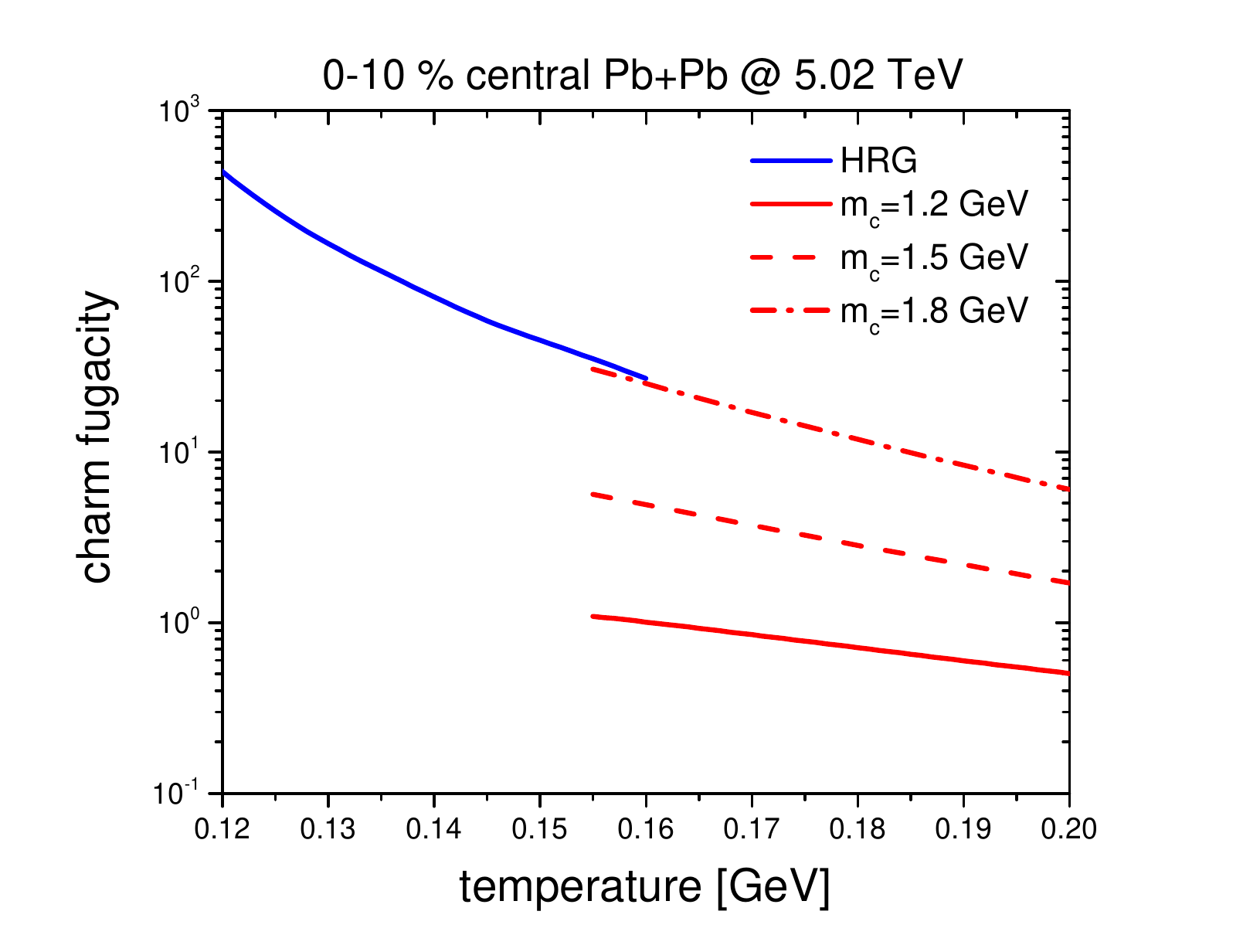}
    \caption{charm fugacity as a function of temperature in 0-10 \% central Pb+Pb collisions at $\sqrt{s_{\rm NN}}=$ 5.02 TeV}
    \label{fugacity-fig}
\end{figure}

Fig.~\ref{fugacity-fig} shows charm fugacity as a function of temperature in 0-10 \% central collisions at $\sqrt{s_{\rm NN}}=$ 5.02 TeV, supposing charm quark number is conserved ($dN_{c\bar{c}}/dy=12.95$).
The total entropy is obtained from the volume (4997 ${\rm fm^3}$) in the statistical model~\cite{Andronic:2021dkw} and the entropy density in the lattice EoS~\cite{Borsanyi:2010cj} and assumed to be constant during the time-evolution.
Then the charm fugacity $g_c$ in hadron gas phase is obtained from 
\begin{eqnarray}
g_c^{HG}=\frac{dN_{c\bar{c}}}{dy}\bigg/
\bigg(V\sum_i \frac{D_i}{2\pi^2}\int dp p^2 
e^{-\sqrt{m_i^2+p^2}/T}\bigg),
\label{fugacity-HG}
\end{eqnarray}
where $V$ is the volume of mid-rapidty, $i$ covers all charm hadrons, $D_i$ is spin degeneracy of charm hadron $i$.
Since a nuclear matter in mid-rapidity at LHC is baryon-free, Eq.~(\ref{fugacity-HG}) does not need baryon chemical potential.
One can see that charm fugacity is around 30 near the phase transition temperature as in the statistical model~\cite{Andronic:2021dkw}, and then increases rapidly with lowering temperature.
In QGP phase, however, the charm fugacity strongly depends on charm quark mass:
\begin{eqnarray}
g_c^{QGP}=\frac{dN_{c\bar{c}}}{dy}\bigg/
\bigg(V \frac{D_c}{2\pi^2}\int dp p^2 
e^{-\sqrt{m_c^2+p^2}/T}\bigg),
\label{fugacity-QGP}
\end{eqnarray}
where $D_c=6$ is charm color-spin degeneracy and $m_c$ charm quark mass.
If $m_c$ is 1.2 GeV as in the QCD Lagrangian, charm fugacity is as low as 1.0 in QGP phase and then suddenly jumps to 30 at $T_c$, which seems not so realistic.
On the other hand, a massive charm quark of around 1.8 GeV smoothly connects charm fugacity in QGP phase to that in hadron gas phase.
This discrepancy of charm fugacity on the QGP side and on the hadron gas side is solved if the charm density, which corresponds to the denominator in Eqs.~(\ref{fugacity-HG}) and (\ref{fugacity-QGP}) excluding the volume $V$, is the same at $T_c$ by adjusting $m_c$ properly. 

In this study we calculate the number density of heavy flavor in hadron gas using the HRG model and the number density of heavy quark in QGP using the quasiparticle model, and then extract the effective mass of heavy quark by matching both of them at $T_c$. 
Since the phase transition is still crossover at a finite baryon chemical potential unless the chemical potential is too large, the same procedure is carried out up to a certain amount of baryon chemical potential and the dependence of heavy quark mass on baryon chemical potential is investigated.

This paper is organized as follows:
In Sec.~\ref{grand-canonical} charm number density is calculated first in hadron gas phase by using the HRG model and then charm quark number density in QGP in grand canonical ensemble. Assuming both of them are equal at $T_c$, the effective mass of charm quark is extracted along the phase boundary with varying baryon chemical potential in grand canonical ensemble.
In Sec.~\ref{canonical} the same calculations are carried out in canonical ensemble which is more realistic for relativistic heavy-ion collisions. And we show the results of bottom quark mass at $T_c$.
Finally, a summary is given in Sec.~\ref{summary}.

\section{Grand canonical ensemble}\label{grand-canonical}
Suppose one prepares thermalized hadron gas (HG) and thermalized QGP and bring both of them to near the temperature for phase transition.
Since the phase transition is crossover for not so a large baryon chemical potential $\mu_B$, charm number densities in HG and in QGP will be same:
\begin{eqnarray}
\sum_M \frac{D_M}{2\pi^2}\int dp p^2 
e^{-\sqrt{m_M^2+p^2}/T_c}\nonumber\\
+\sum_B \frac{D_B}{2\pi^2}\int dp p^2 
e^{-(\sqrt{m_B^2+p^2}-\mu_B)/T_c}\nonumber\\
=\frac{3}{\pi^2}\int dp p^2 
e^{-(\sqrt{m_c^2+p^2}-\mu_B/3)/T_c},
\label{charm}
\end{eqnarray}
where $M$ and $B$ respectively indicate charm meson and charm baryon in hadron gas and $D_i$ and $m_i$ are their spin degeneracy and mass, respectively.
In principle mesons should follow the Bose-Einstein distribution and baryons the Dirac-Fermi distribution.
But charm meson and baryon are heavy enough around $T_c$ to be substituted with the simple Boltzmann distribution.

We invoke the hadron resonance gas model on the left side and take into account all charm hadrons up to $D_3^*(2750)$ for $D$ meson, up to $D_{s1}^*(2860)^\pm$ for $D_s$, up to $\Lambda_c(2940)^+$ for $\Lambda_c$, up to $\Sigma_c(2520)$ for $\Sigma_c$, up to $\Xi_c(2970)$ for $\Xi_c$ and up to 
$\Omega_c(2770)^0$ for $\Omega_c$ from the particle data group~\cite{Workman:2022ynf}.
Though there are more higher excited states, their spin degeneracies are not confirmed yet and so they are excluded.
We do not also include charmonia, because their contribution is very little at $T_c$.
On the right hand side $m_c$ is effective charm quark mass in QGP.

In grand canonical ensemble one can assume $\mu_u=\mu_d=\mu_s=\mu_c=\mu_B/3$, because three quarks form a baryon.
The chemical reaction between three quarks and a baryon, $q_i+q_j+q_k \leftrightarrow B_a$, satisfies the following relation of chemical potentials:
\begin{eqnarray}
\mu_i+\mu_j+\mu_k=\mu_{B_a}
\label{chemical}
\end{eqnarray}
where $i,j,k$ are quark flavors and $a$ is baryon species.
One can handle the relative abundance of a specific flavor by changing the corresponding quark chemical potential but still satisfying Eq.~(\ref{chemical}).
In this way one can control isospin density, strangeness density, electric charge density of a matter.
In other words, there are many different kinds of baryonic nuclear matter.
However, even a nuclear matter produced in heavy-ion collisions is not free of electric charge and isospin asymmetric, because colliding nuclei have many neutrons than protons.
We therefore use a simple case ($\mu_u=\mu_d=\mu_s=\mu_c=\mu_B/3$) in this study.

The number density of anticharm quark is obtained in the same way as in Eq.~(\ref{charm}):  
\begin{eqnarray}
\sum_{\overline{M}} \frac{D_{\overline{M}}}{2\pi^2}\int dp p^2 
e^{-\sqrt{m_{\overline{M}}^2+p^2}/T_c}\nonumber\\
+\sum_{\overline{B}} \frac{D_{\overline{B}}}{2\pi^2}\int dp p^2 
e^{-(\sqrt{m_{\overline{B}}^2+p^2}+\mu_B)/T_c}\nonumber\\
=\frac{3}{\pi^2}\int dp p^2 
e^{-(\sqrt{m_{\overline{c}}^2+p^2}+\mu_B/3)/T_c},
\label{anticharm}
\end{eqnarray}
where $\overline{M}$ and $\overline{B}$ respectively indicate anticharm meson and anticharm baryon in hadron gas, and $m_{\overline{c}}$ anticharm quark mass in QGP.
We note that the sign of baryon chemical potential $\mu_B$ is opposite in Eq.~(\ref{anticharm}).

The critical temperature depends on $\mu_B$, and can be parameterized as~\cite{Cassing:2007nb,Moreau:2019vhw}
\begin{eqnarray}
T_c(\mu_B)=T_c(0)\sqrt{1-\alpha \mu_B^2}
\label{tc}
\end{eqnarray}
where $T_c(0)=$ 0.158 GeV and $\alpha=$ 0.974, which is valid up to $\mu_B=$ 0.5 GeV.
Applying Eq.~(\ref{tc}) to Eqs.~(\ref{charm}) and (\ref{anticharm}), one can obtain (anti)charm quark mass for a finite $\mu_B$ along the phase boundary.

\begin{figure} [b!]
\includegraphics[width=8.5 cm]{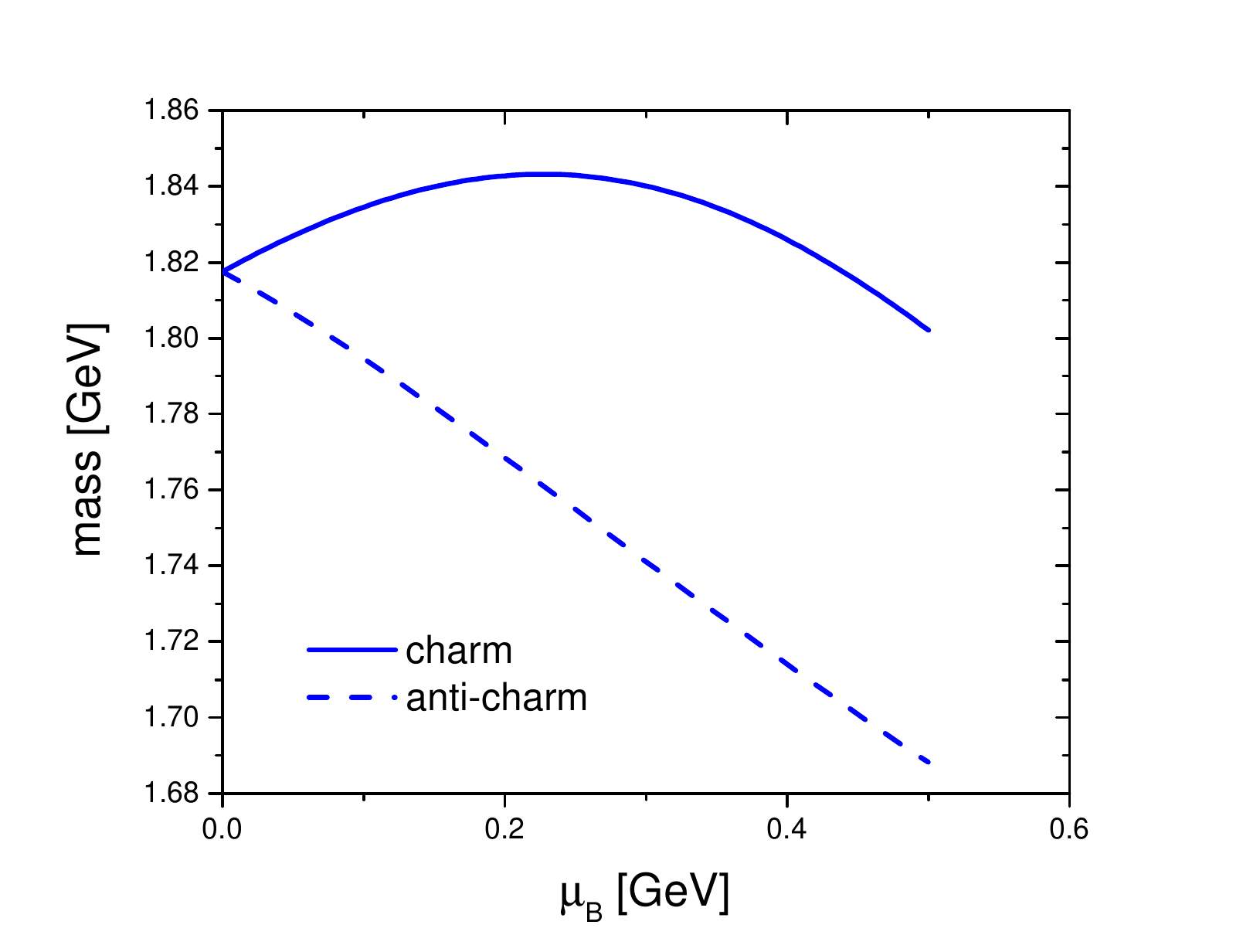}
    \caption{charm and anticharm quark masses as a function of baryon chemical potential at the phase transition in grand canonical ensemble}
    \label{grand-fig}
\end{figure}

Fig.~\ref{grand-fig} shows (anti)charm quark mass as a function of baryon chemical potential in grand canonical ensemble.
Eqs.~(\ref{charm}) and (\ref{anticharm}) are numerically solved by using the Newton's method.
One can see that the effective mass of (anti)charm quark is about 1.82 GeV at $\mu_B$=0, which is much larger than charm quark bare mass in QCD Lagrangian but close to $D$ meson mass which is the lowest state of charm meson.
We note that the large mass of charm quark is consistent with the results from our recent study in heavy-ion collisions~\cite{Song:2024hvv}.

As baryon chemical potential increases, 
charm quark mass a little increases and then decreases, while anticharm quark mass continually decreases.
One can easily understand the different behaviors from Eqs.~(\ref{charm}) and (\ref{anticharm}).
At low baryon chemical potential, charm meson density is still dominant over charm baryon density and Eq.~(\ref{charm}) is approximated into 
\begin{eqnarray}
\sum_M \frac{D_M}{2\pi^2}\int dp p^2 
e^{-\sqrt{m_M^2+p^2}/T_c}\nonumber\\
\approx\frac{3}{\pi^2}\int dp p^2 
e^{-(\sqrt{m_c^2+p^2}-\mu_B/3)/T_c}.
\label{charm-approx}
\end{eqnarray}
If the left hand side is constant, because it does not depend on $\mu_B$, $m_c$ in the right hand side should increase with increasing $\mu_B$ for equality.
However, it is opposite in Eq.~(\ref{anticharm}), because the sign of $\mu_B$ is positive.

Once baryon chemical potential exceeds some limit, charm baryon density is more dominant than charm meson density and Eq.~(\ref{charm}) turns to 
\begin{eqnarray}
\sum_B \frac{D_B}{2\pi^2}\int dp p^2 
e^{-(\sqrt{m_B^2+p^2}-\mu_B)/T_c}\nonumber\\
\approx\frac{3}{\pi^2}\int dp p^2 
e^{-(\sqrt{m_c^2+p^2}-\mu_B/3)/T_c}
\label{charm-approx2a}
\end{eqnarray}
or
\begin{eqnarray}
\sum_B \frac{D_B}{2\pi^2}\int dp p^2 
e^{-\sqrt{m_B^2+p^2}/T_c}\nonumber\\
\approx\frac{3}{\pi^2}\int dp p^2 
e^{-(\sqrt{m_c^2+p^2}+2\mu_B/3)/T_c}.
\label{charm-approx2b}
\end{eqnarray}
Therefore, as $\mu_B$ increases, $m_c$ should decrease. 

As for anticharm quark anticharm meson density is always dominant over anticharm baryon density, because the anticharm baryon density is strongly suppressed with increasing $\mu_B$.
That is why anticharm quark mass continually decreases with increasing $\mu_B$.

\section{Canonical ensemble}\label{canonical}
A hot and dense matter produced in heavy-ion collisions does not have net charm flavor, because charm is always produced by pair~\cite{Andronic:2021erx}.
However, in the grand canonical ensemble, charm number density is larger or smaller than the number density of anticharm, depending on the baryon chemical potential.

To be more realistic, charm flavor in heavy ion collisions should be treated in canonical ensemble rather than in grand canonical ensemble~\cite{Rafelski:1980gk}.
Then Eq.~(\ref{charm}) for charm is modified into 
\begin{eqnarray}
\sum_M \frac{D_M}{2\pi^2}\int dp p^2 
e^{-(\sqrt{m_M^2+p^2}+\mu_B/3-\mu_c)/T_c}\nonumber\\
+\sum_B \frac{D_B}{2\pi^2}\int dp p^2 
e^{-(\sqrt{m_B^2+p^2}-2\mu_B/3-\mu_c)/T_c}\nonumber\\
=\frac{3}{\pi^2}\int dp p^2 
e^{-(\sqrt{m_c^2+p^2}-\mu_c)/T_c}=n_c,
\label{charm2}
\end{eqnarray}
and Eq.~(\ref{anticharm}) for anticharm into 
\begin{eqnarray}
\sum_{\overline{M}} \frac{D_{\overline{M}}}{2\pi^2}\int dp p^2 
e^{-\sqrt{(m_{\overline{M}}^2+p^2}-\mu_B/3-\mu_{\overline{c}})/T_c}\nonumber\\
+\sum_{\overline{B}} \frac{D_{\overline{B}}}{2\pi^2}\int dp p^2 
e^{-(\sqrt{m_{\overline{B}}^2+p^2}+2\mu_B/3-\mu_{\overline{c}})/T_c}\nonumber\\
=\frac{3}{\pi^2}\int dp p^2 
e^{-(\sqrt{m_{\overline{c}}^2+p^2}-\mu_{\overline{c}})/T_c}=n_{\overline{c}},
\label{anticharm2}
\end{eqnarray}
where charm and anticharm number densities are separately controlled by $\mu_c$ and by $\mu_{\overline{c}}$.
Here the baryon chemical potential $\mu_B$ is contributed from up, down and strange flavors.

Comparing Eq.~(\ref{charm}) and Eq.~(\ref{charm2}), one can see that both equations are equivalent to each other, if $e^{(\mu_B/3-\mu_c)/T_c}$ is multiplied to the both sides of Eq.~(\ref{charm2}).
Eq.~(\ref{anticharm}) and Eq.~(\ref{anticharm2}) are also equivalent, $e^{-(\mu_B/3+\mu_{\overline{c}})/T_c}$ multiplied to the both sides of Eq.~(\ref{anticharm2}).
Therefore, charm and anticharm quark masses in canonical ensemble are same as in grand canonical ensemble in Fig.~\ref{grand-fig}.

As next step, we can find $\mu_c$ and $\mu_{\overline{c}}$ which vanishes net charm density, that is, $n_c=n_{\overline{c}}$.
For this purpose we redefine $\mu_c$ and $\mu_{\overline{c}}$ as following:
\begin{eqnarray}
\mu_c^S&=&\frac{\mu_c+\mu_{\overline{c}}}{2},\nonumber\\
\mu_c^A&=&\frac{\mu_c-\mu_{\overline{c}}}{2},
\end{eqnarray}
such that
\begin{eqnarray}
\mu_c&=&\mu_c^S+\mu_c^A,\nonumber\\
\mu_{\overline{c}}&=&\mu_c^S-\mu_c^A.
\label{muc-new}
\end{eqnarray}

Substituting Eq.~(\ref{muc-new}) into Eqs.~(\ref{charm2}) and (\ref{anticharm2}),  
one finds $\mu_c^S$ controls the number of charm quark pairs, which is related to charm fugacity in the statistical model~\cite{Andronic:2021erx}, while $\mu_c^A$ controls relative abundance of charm to anticharm.
Since $\mu_c^S$ is an overall factor, we can neglect it and focus on $\mu_c^A$.
The $\mu_c^A$ which realizes vanishing net charm density is given by
\begin{eqnarray}
\mu_c^A=\frac{T_c}{2}\ln\bigg(\int dp p^2 
e^{-\sqrt{m_c^2+p^2}/T_c}\nonumber\\
\bigg/ \int dp p^2 
e^{-\sqrt{m_{\overline{c}}^2+p^2}/T_c}\bigg).
\label{muc}
\end{eqnarray}

Taking a nonrelativistic approximation, $\sqrt{m_c^2+p^2}\approx m_c+p^2/2m_c$, Eq.~(\ref{muc}) is simplified into 
\begin{eqnarray}
\mu_c^A\approx \frac{m_{\overline{c}}-m_c}{2}+\frac{3T_c}{4}\ln\frac{m_c}{m_{\overline{c}}}.
\label{muc-approx}
\end{eqnarray}

\begin{figure} [h!]
    \includegraphics[width=8.5 cm]{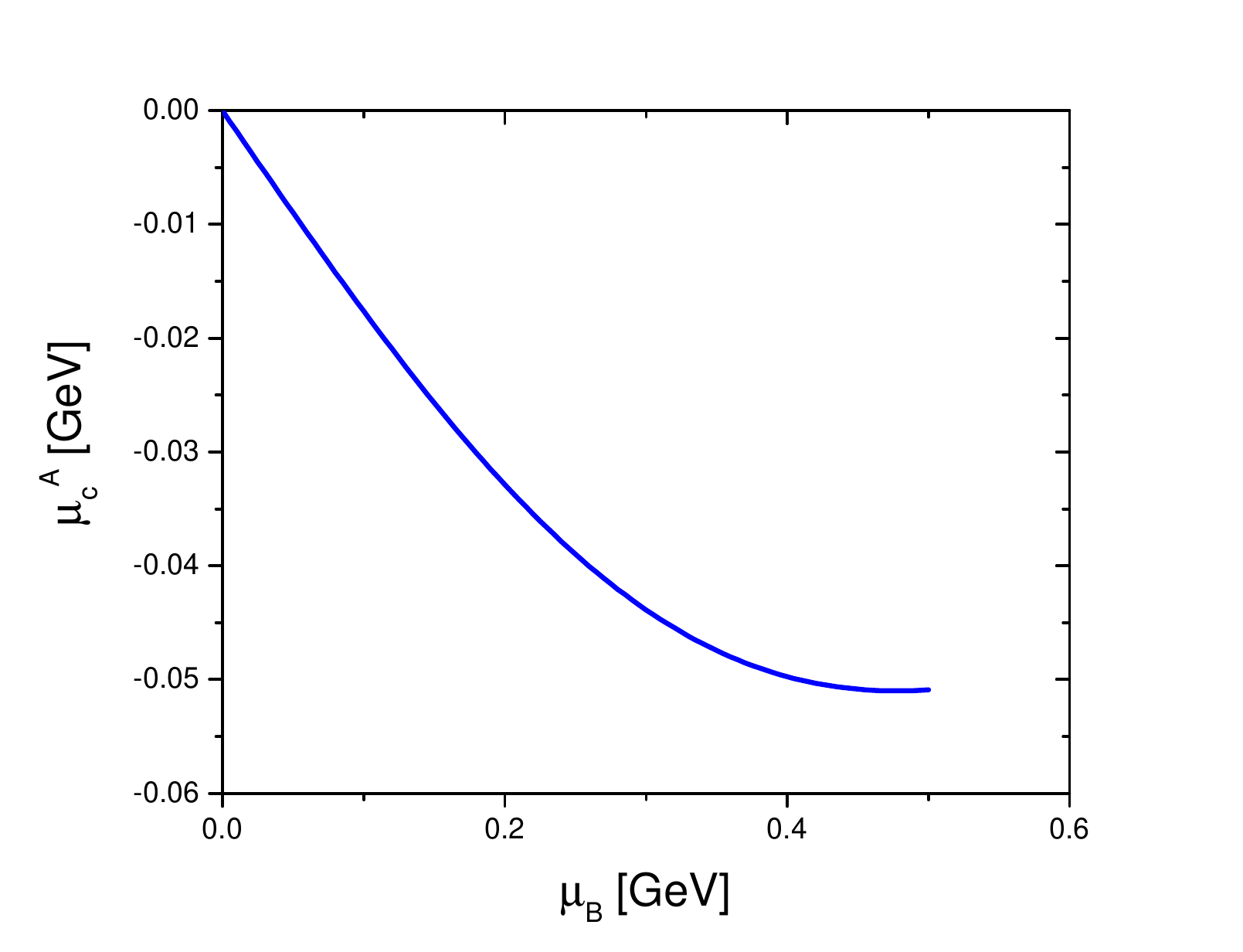}
    \caption{asymmetric charm chemical potential for vanishing net charm density as a function of baryon chemical potential at the phase transition in canonical ensemble}
    \label{canonical-fig}
\end{figure}

Fig.~\ref{canonical-fig} shows $\mu_c^A$ which vanishes net charm density in matter as a function of baryon chemical potential that is contributed from up, down and strange flavors.
As shown in Eqs.~(\ref{muc}) and (\ref{muc-approx}), the asymmetric charm chemical potential starts with 0 at $\mu_B=0$ where charm mass and anticharm mass are same, and then decreases with increasing $\mu_B$.
However, it saturates at large $\mu_B$, because mass difference between charm and anticharm quarks becomes constant as shown in Fig.~\ref{grand-fig}.

One can do the same thing for bottom quark.
On hadronic side all $B$ mesons are included up to $B_2^*(5747)$, all $B_s$ mesons up to $B_{s2}^*(5840)$, all $\Lambda_b$ up to $\Lambda_b(6152)^0$, all $\Sigma_b$ up to $\Sigma_b^*$, all $\Xi_b$ up to $\Xi_b(6100)$ and $\Omega_b^-$.

\begin{figure} [b!]
    \includegraphics[width=8.5 cm]{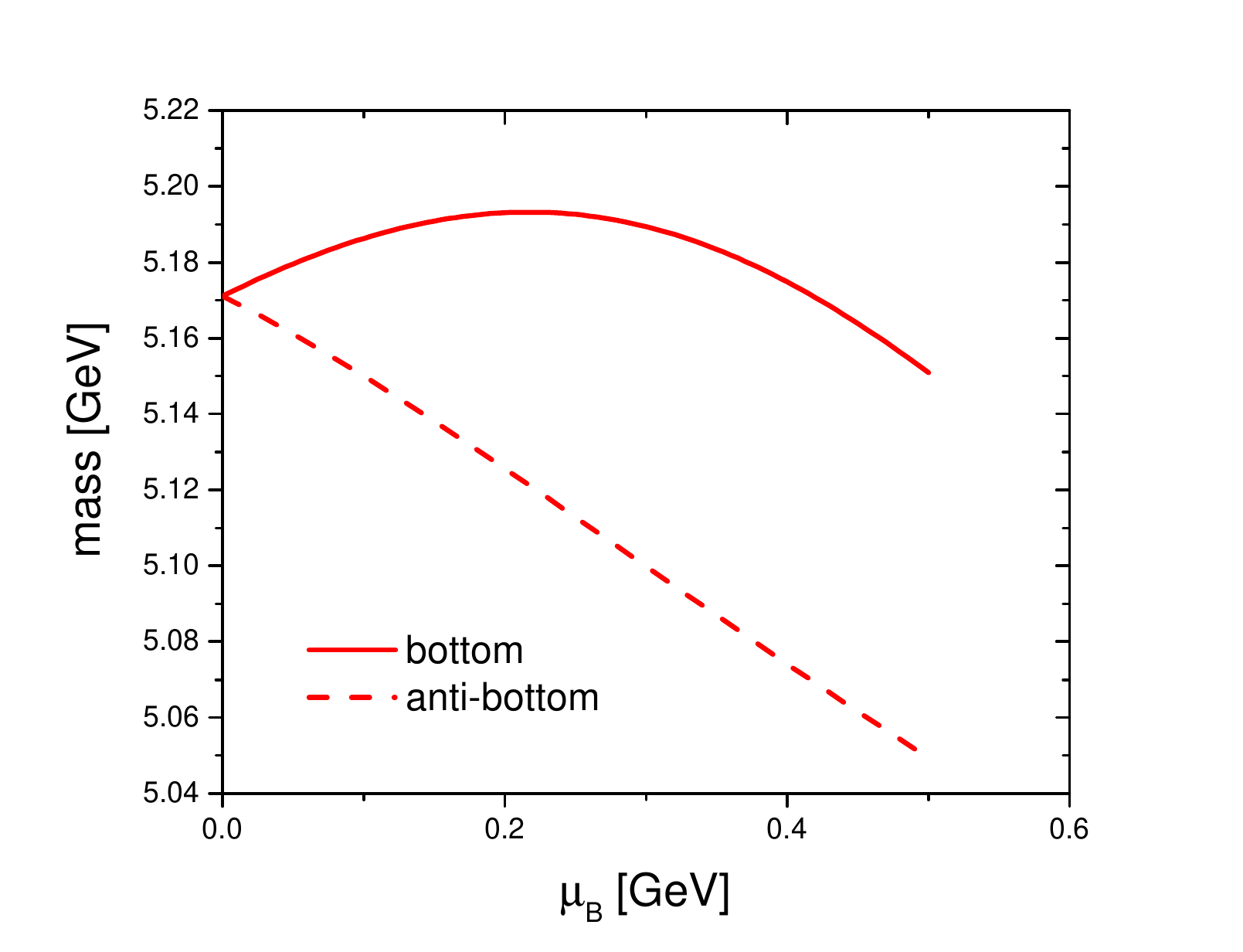}
    \includegraphics[width=8.5 cm]{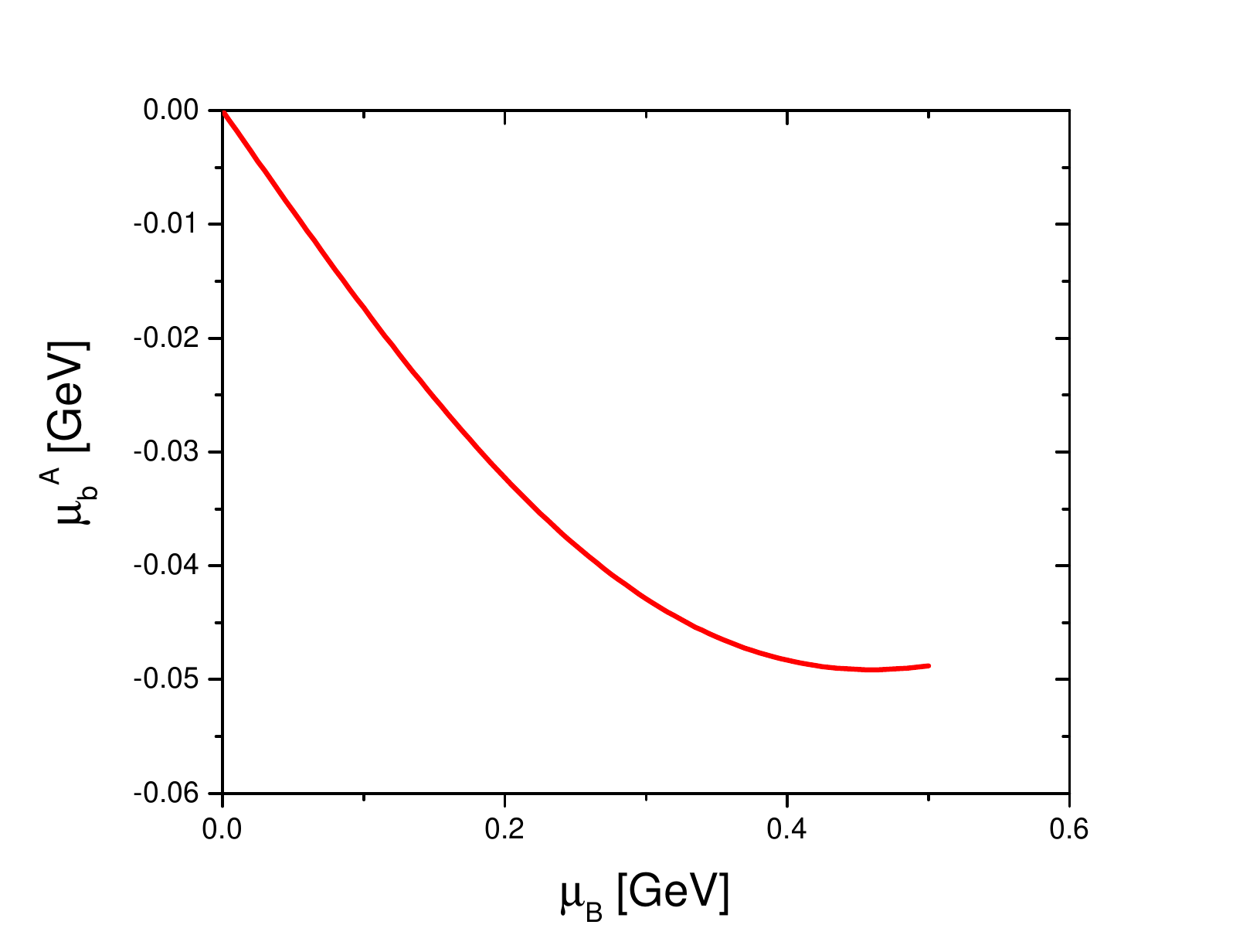}    
    \caption{(upper) bottom and antibottom quark masses and (lower) asymmetric bottom chemical potential for vanishing net bottom density both as a function of baryon chemical potential at the phase transition.}
    \label{grand-bottom}
\end{figure}

We show in Fig.~\ref{grand-bottom} bottom and antibottom quark masses as a function of baryon chemical potential at the phase transition.
Same as charm quark, bottom quark mass slightly increases and then decreases with increasing baryon chemical potential, while antibottom quark mass constantly decreases.

The asymmetric bottom chemical potential which is in the lower panel shows a similar behavior as that of the asymmetric charm chemical potential, because the mass difference between bottom and antibottom quarks is similar to that of charm and anticharm quarks, which is around 100 MeV.

\begin{figure} [h!]
    \includegraphics[width=8.5 cm]{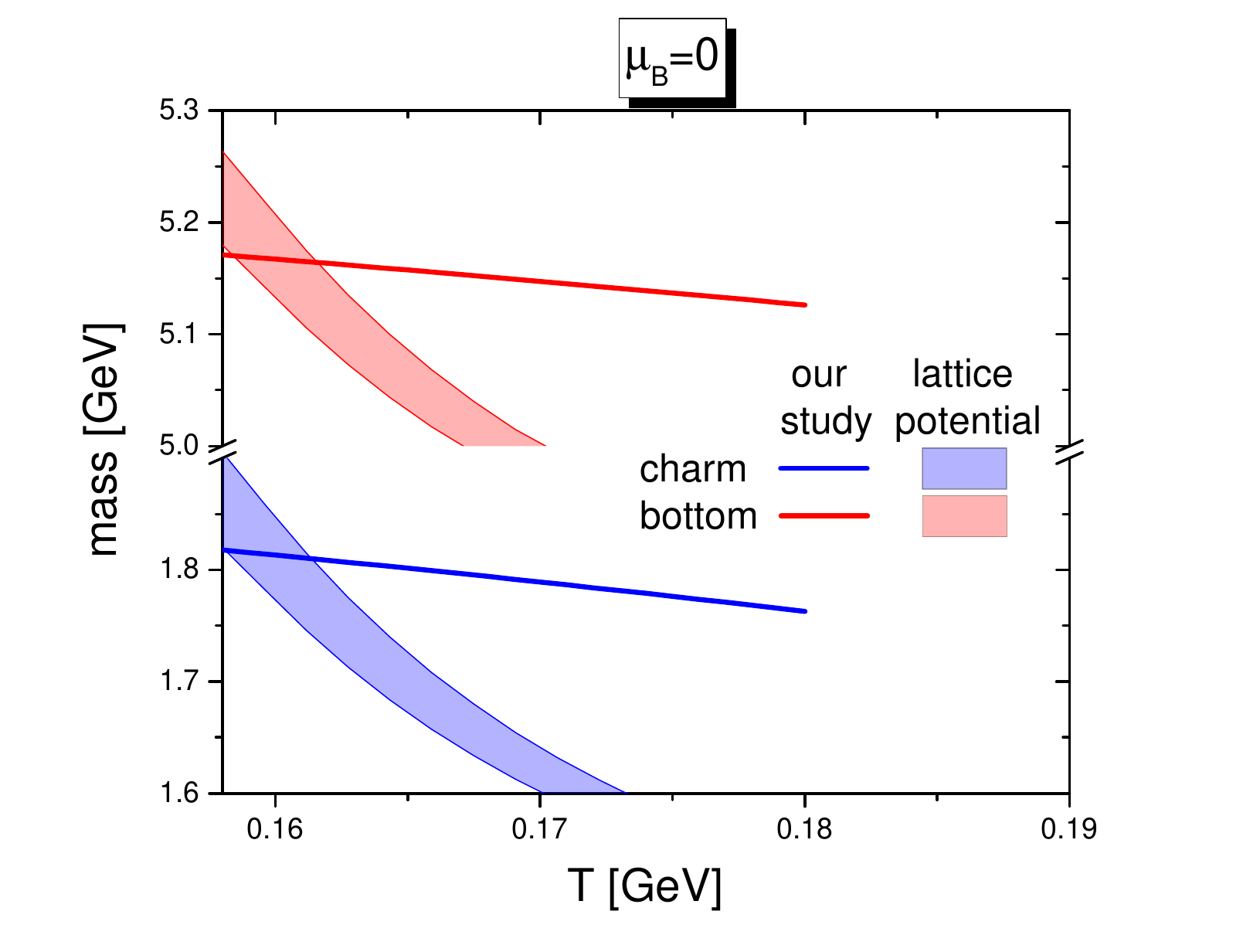}  
    \caption{charm and bottom quark masses as a function of temperature at $\mu_B$=0 in comparison with those from lattice heavy quark potential.}
    \label{mass-tem}
\end{figure}

Finally we discuss the change of charm and bottom quark masses with varying temperature.
We assume that the HRG model works not only at $T_c$ but also a bit further above $T_c$ as shown in the lattice QCD calculations~\cite{Borsanyi:2010cj}.

Fig.~\ref{mass-tem} displays the mass of charm quark and that of bottom quark as a function of temperature at $\mu_B$=0.
Because the HRG model will eventually fail at high temperature,  
the results do not go to too high temperature but end around $T=$ 180 MeV.
One can see that both charm and bottom quark masses decrease with temperature, which is consistent with the behavior of heavy quark mass of several groups who study quarkonium in QGP~\cite{Andronic:2024oxz}.
The reason for the decreasing mass is that, as temperature increases, the excited states of heavy flavor in the HRG model contributes more and the effective mass of heavy quark lowers as a result.

In order to test our results quantitatively, we compare our results with heavy quark mass from the heavy quark lattice potential~\cite{Kaczmarek:2003ph}.
As explained in Ref.~\cite{Gubler:2020hft}, heavy quark mass in QGP is related to the heavy quark potential at infinity, because a heavy quark pair separated infinitely from each other is considered as two open heavy flavors which cannot affect each other:
\begin{eqnarray}
m(T) =m_0+\frac{1}{2}V\bigg(r=\infty,T\bigg),
\label{dmass}
\end{eqnarray}
where $m_0$ is heavy quark bare mass, 1.26 GeV for charm and 4.62 GeV for bottom such that the Schr\"odinger equation with the heavy quark potential can reproduce the vacuum mass of quarkonium ground state~\cite{Andronic:2024oxz}. 
Comparing with $D$ meson mass near $T_c$ from a hadronic effective model~\cite{Montana:2020lfi} and from QCD sum rule at finite temperature~\cite{Gubler:2020hft}, the best fit is obtained from the combination of 15 \% free energy and 85 \% internal energy of heavy quark pair in the former model and 20 \% free energy and 80 \% internal energy in the latter model, which is shown as the lower and upper limits of the colored bands in Fig.~\ref{mass-tem}, respectively.
One can find that the results are consistent with the colored lines near $T_c$ which are obtained by combining the hadron resonance gas model and quasi-particle model.

\section{Summary}\label{summary}

Heavy flavor is one of important probe particles searching for the properties of a QGP produced in relativistic heavy-ion collisions.
It is found that a QGP near the phase transition is not a free gas but a strongly interacting matter like fluid.
Therefore, a parton in QGP gains thermal mass and width through interactions in matter.
The effective mass and/or width of light partons can be extracted from the equation-of-state of QGP which is provided by lattice QCD calculations, but heavy quark mass cannot be constrained by it.

In this study, using that the phase transition between QGP and hadron gas is crossover at small baryon chemical potential, we assume that the number density of heavy quark and that of heavy antiquark are separately conserved during the phase transition at $T_c$.
For QGP phase we simply use a Boltzmann distribution of charm or bottom quarks with its effective mass and for hadron gas phase the hadron resonance gas model is utilized, where all hadron states with a definite spin number in the particle data book are considered.
As a result, the effective mass of (anti)charm quark is about 1.82 GeV and that of (anti)bottom quark about 5.17 GeV at vanishing baryon chemical potential.

As baryon chemical potential increases, heavy quark mass and the mass of heavy antiquark are split.
Heavy quark mass slightly increases and then decreases with increasing baryon chemical potential.
On the other hand, the mass of heavy antiquark constantly decreases.

Assuming that the hadron resonance gas model works beyond the critical temperature, though it will eventually break down at high temperature, we have found that the mass of charm quark and that of bottom quark decrease with temperature.
This behavior is consistent with the assumption of several groups who work on quarkonium in QGP.

Heavy quark mass in QGP is an important quantity for studying quarkonium production and dissociation in heavy-ion collisions.
The results in this study will provide useful information for the quarkonium study not only in high energy collisions which correspond to small baryon chemical potential but also in  intermediate energy collisions that is relevant to relatively large baryon chemical potential.

\section*{Acknowledgements}
The authors acknowledge inspiring discussions with E. Bratkovskaya and J. Aichelin.
Furthermore, we acknowledge support by the Deutsche Forschungsgemeinschaft (DFG, German Research Foundation) through the grant CRC-TR 211 'Strong-interaction matter under extreme conditions' - Project number 315477589 - TRR 211. 
This work is also supported by the China Scholarship Council under Contract No.~202306770010.
The computational resources have been provided by the LOEWE-Center for Scientific Computing and the "Green Cube" at GSI, Darmstadt and by the Center for Scientific Computing (CSC) of the Goethe University, Frankfurt.

\bibliography{main}

\end{document}